\documentclass[floatfix,twocolumn,showpacs,floats,aps,superscriptaddress]{revtex4-1}
\usepackage[pdftex]{graphicx}
\usepackage{amsmath}
\usepackage{amssymb}
\usepackage{bm}

\usepackage[normalem]{ulem}
\usepackage{color}

\begin{document}
\title{Regular packings on periodic lattices}

\author{Tadeus Ras}
\altaffiliation[Present address: ]{Fachbereich Physik, Universit\"at Konstanz, 78457~Konstanz, Germany}
\author{Rolf Schilling}
\affiliation{Institut f\"ur Physik, Johannes Gutenberg-Universit\"at Mainz,
  Staudinger Weg 7, D-55099 Mainz, Germany}

\author{Martin Weigel}
\affiliation{Institut f\"ur Physik, Johannes Gutenberg-Universit\"at Mainz,
  Staudinger Weg 7, D-55099 Mainz, Germany}
\affiliation{Applied Mathematics Research Centre, Coventry University,
  Coventry, CV1~5FB, England}

\date{\today}
%\maketitle

\begin{abstract}
  We investigate the problem of packing identical hard objects on regular lattices in
  $d$ dimensions. Restricting configuration space to parallel alignment of the
  objects, we study the densest packing at a given aspect ratio $X$. For rectangles
  and ellipses on the square lattice as well as for biaxial ellipsoids on a simple
  cubic lattice, we calculate the maximum packing fraction $\varphi_d(X)$. It is
  proved to be continuous with an infinite number of singular points $X^{\rm
    min}_\nu$, $X^{\rm max}_\nu$ $\nu=0$, $\pm 1$, $\pm 2$, $\ldots$. In two
  dimensions, all maxima have the same height, whereas there is a unique global
  maximum for the case of ellipsoids. The form of $\varphi_d(X)$ is discussed in the
  context of geometrical frustration effects, transitions in the contact numbers and
  number theoretical properties. Implications and generalizations for more general
  packing problems are outlined.
\end{abstract}

%b) Electronic mail: rolf.schilling@uni-mainz.de\\
%c) Permanent address\\

%\newpage
\maketitle

The question of how densely objects can fill a volume has attracted both,
mathematicians and physicists, for centuries. One famous problem is that of packing
spheres. In 1611, Kepler conjectured that the face-centered cubic (fcc) and hexagonal
close packed (hcp) lattice configurations of identical spheres yield the highest
packing fraction $\varphi_{\rm max}^{d=3}=\pi/\sqrt{18} \cong 0.7404$. Gau{\ss} could
show in 1831 that these are the optimal {\em periodic\/} packings of spheres, but
only very recently it was proved that they are optimal within all possible
arrangements \cite{1}. Even for disks in the plane, the corresponding proof of
optimality of the hexagonal packing with $\varphi_{\rm max}^{d=2} =\pi/\sqrt{12}
\cong 0.9069$ was only found in 1943 \cite{1.5}.
%While the structures and densities of lattice packings for hyper-spheres are known up
%to dimension $d=8$, no rigorous results are available for the absolute densest
%packings of spheres beyond three dimensions (3d) \cite{3}.
Apart from its theoretical attraction along with its relation to coding theory
\cite{3}, packing is a problem of practical relevance. Not only have practitioners
long known that a densest packing of oranges or cannon balls can be achieved via
hexagonal layering but, more recently, packing problems have received substantial
attention in engineering and operations research as problems of optimizing yields in
production or minimizing leakage currents in integrated circuits (see, e.g.,
Ref.~\cite{1.8}).

In physics, periodic packings \cite{2,3,4} are relevant for describing and
understanding crystalline materials. In contrast, random close packings \cite{5},
i.e., maxima of the packing fraction under some local dynamics starting from loosely
packed configurations, have been used to model glasses \cite{6} and granular
materials \cite{7}. For spheres in 3d, random close packing leads to a packing
fraction $\varphi_{\rm RCP}^{d=3} \approx 0.64$, significantly below $\varphi_{\rm
  max}^{d=3}$. The hard objects considered in such packings need not be spheres, but
can be more general convex bodies. Although recently there has been extensive
numerical work using techniques from dynamic programming and heuristic optimization,
complemented by experiments, for studying periodic packings \cite{7.5} or random
close packing \cite{7.7} for non-spherical objects, there is a lack of analytical
understanding of these problems. For random close packing, it has been observed that
the packing fraction increases over $\varphi_{\rm RCP}^{d=3} \approx 0.64$ as spheres
are replaced by ellipsoids, and might even approach $\varphi_{\rm
  max}^{d=3}=\pi/\sqrt{18} \cong 0.7404$ in some cases \cite{7.7}. Concerning
periodic packings, an affine transformation maps the fcc/hcp sphere packing to a
periodic lattice packing of identically aligned ellipsoids with maximum packing
fraction $\varphi_{\rm max}^{d=3}=\pi/\sqrt{18}$. That non-parallel arrangement of
ellipsoids of revolution may exceed $\pi/\sqrt{18}$ has been predicted in
Refs.~\cite{8,9}. Such super-dense packings of ellipsoids were studied recently in
more detail \cite{10,11}. Particularly, it has been shown that $\varphi\cong0.7707$
for all aspect ratios $X \geq \sqrt{3}$ \cite{10}.

We make progress in the analytical understanding of the problem of packings of
non-spherical bodies by taking a complementary approach. Instead of finding the
lattice structure that maximizes the packing fraction for a given type ${\cal K}$ of
objects, we start out from a fixed Bravais lattice $\Lambda$ and attach a body ${\cal
  K}$ of the same shape and orientation ${\boldsymbol \omega}$ to each lattice site
(at its center of mass, say). We then determine the maximum packing fraction as a
function of ${\cal K}$, i.e., as a function of the parameters characterizing its
shape and orientation. To the best of our knowledge, this problem has not been
studied before. Our approach may contribute to describing, for instance, plastic
crystals, i.e., lattices with a molecule fixed at each site. In particular, aromatic
molecules can be approximately described by hard ellipsoids. Similarly, applications
are envisaged in operations research and manufacturing. Finally, insight into the
frustration effects generated by the competing length scales of ${\cal K}$ and
$\Lambda$ could contribute to the understanding of packings without a pre-determined
lattice structure.

Consider a class of identical $d$-dimensional convex bodies ${\cal K}$ whose shape
depends merely on their ``length'' $l$ and ``width'' $w$, and consequently are
characterized by a single parameter $X=l/w$, the aspect ratio. As a general example
one might think of a $d$-dimensional ellipsoid of revolution. Fixing the aspect ratio
$X$ and orientation ${\boldsymbol \omega}$, proportional rescaling of the bodies
allows to reach the maximum packing fraction without overlaps, $\varphi_d (X,
{\boldsymbol \omega})$. This fraction varies with ${\boldsymbol \omega}$, and we are
interested in the maximum packing fraction irrespective of orientation, $\varphi_d(X)
= \max_{\boldsymbol \omega} \varphi_d (X, {\boldsymbol \omega})$. The maximum
$\varphi_d(X)$ is continuous as a function of $X$. Here, we only outline the idea of
the rigorous proof \cite{13}. Let us assume that $ \varphi_d(X) $ is discontinuous at
some $ X_0=l_0/w_0 $ where, e.g. it jumps from $ \varphi_- $ to $ \varphi_+>\varphi_-
$, with $ \varphi_\pm=\lim\limits_{\varepsilon\to 0} \varphi_d(X_0\pm\epsilon)$. The
convex bodies at $ \varphi_- $ and $ \varphi_+ $ are characterized by $ (l_-,w_-) $
and $ (l_+,w_+) $, respectively. Both pairs differ from each other, as $
\varphi_-\neq\varphi_+ $. Of course, it is $ l_+/w_+=l_-/w_-=X_0 $. Now, starting
from the configuration at $ \varphi_+ $, we {\em continuously} decrease the length of
the hard objects. Consequently, without change of orientation, both the aspect ratio
$ X $ and the corresponding packing fraction $ \tilde{\varphi}_d(X) $ decrease
continuously from $ X_0 $ and $ \varphi_+=\tilde{\varphi}_d(X_0) $,
respectively. Below but arbitrarily close to $ X_0 $, $ \tilde{\varphi}_d(X) $ must
be arbitrarily close to $ \varphi_+ $, due to its continuity. On the other hand, it
is $ \varphi_d(X)\geq\tilde{\varphi}_d(X) $ for all $ X\leq X_0 $, since $
\varphi_d(X) $ is the {\em maximum} packing fraction by definition. Therefore, even
if $ \varphi_d(X)=\tilde{\varphi}_d(X) $ holds (instead of $ \geq $) for all $ X\leq
X_0 $, we get $ \varphi_-=\lim\limits_{\varepsilon\to 0}
\varphi_d(X_0-\epsilon)=\varphi_+ $. This contradicts the original assumption $
\varphi_+>\varphi_- $. Consequently, $ \varphi_d $ must be continuous.

\begin{figure}
\includegraphics[width=\linewidth]{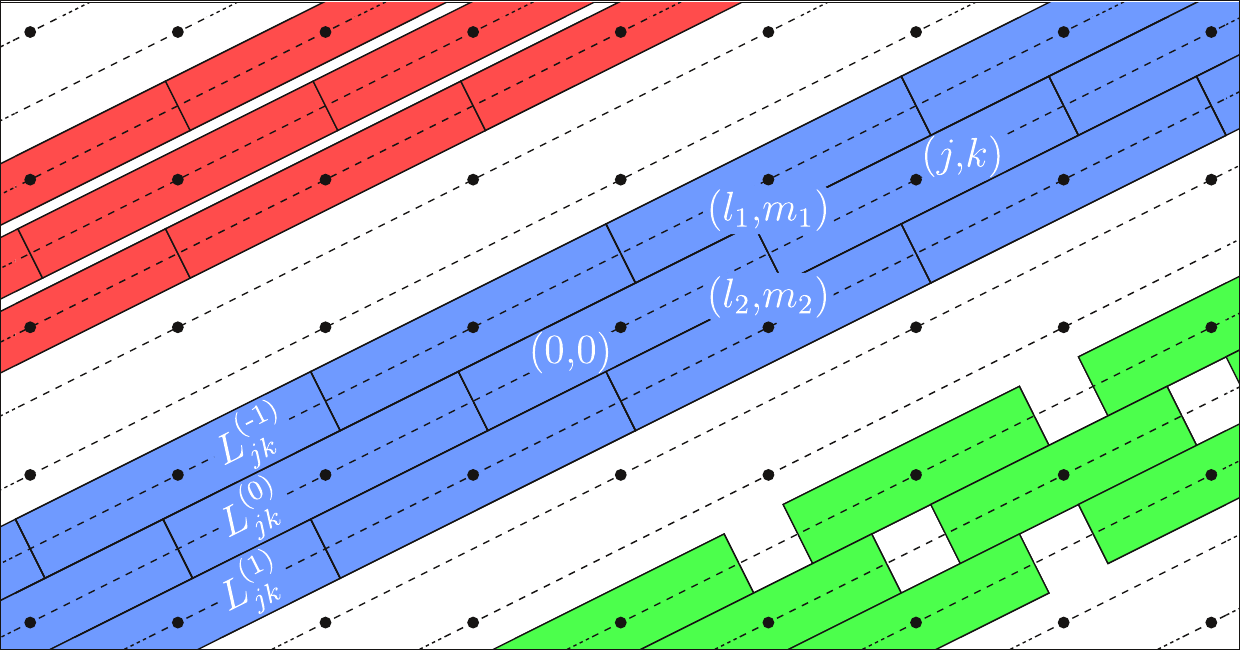}
\caption{(Color online) Maximum packing configurations of rectangles with aspect
  ratios $X=4$ (green, bottom), $X=X_{21}^{\text{max}}=5$ (blue, middle), and $X=7$
  (red, top), respectively. Lattice lines $L_{jk}^{(\sigma)}$, $\sigma=0$, $\pm 1$,
  $\ldots$ for $(j,k)=(2,1)$ are indicated with dashed lines.}
\label{fig:1}
\end{figure}

\begin{figure}
\includegraphics[width=\linewidth]{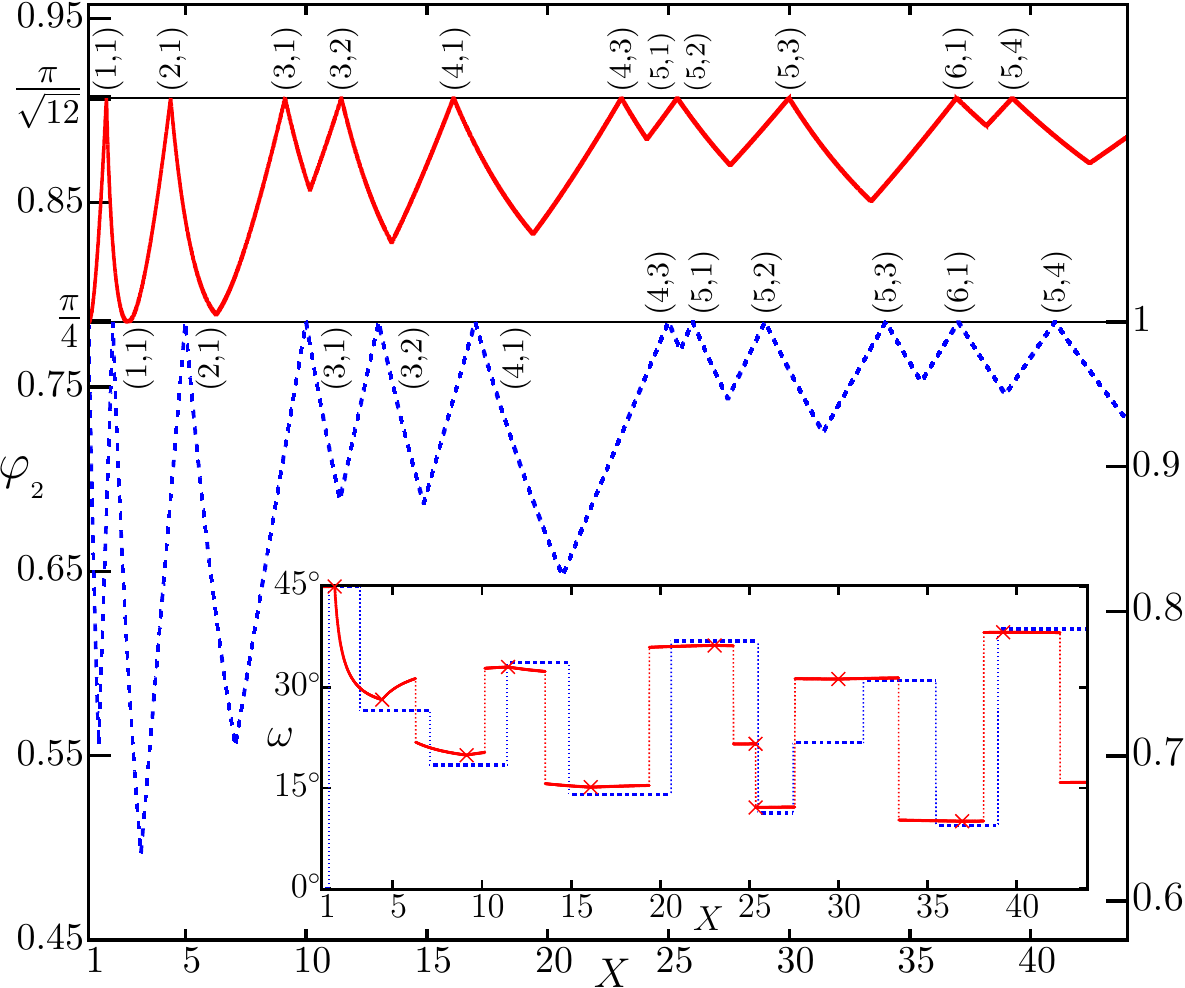}
\caption{(Color online) Maximum packing fraction $\varphi_2(X)$ for rectangles
  (dashed line, right scale) and for ellipses (solid line, left scale). Inset:
  orientation angle $\omega$ as a function of $X$ for rectangles (dashed line) and
  ellipses (solid line). The crosses correspond to the positions of the maxima of
  $\varphi_2(X)$ for ellipses.  }
\label{fig:2}
\end{figure}

We now turn to the calculation of $\varphi_d(X)$ for specific hard objects. As an
example in two dimensions (2d), we study a square lattice with lattice constant
$a=1$.  Consider first the case of rectangles of length $l$ and width $w$. Imagine
two identical rectangles with common direction ${\bf e}=(\cos\omega, \sin\omega)$, of
their long side, attached with their centers to lattice sites $(0,0)$ and ${\bf
  R}_{jk}=(j,k)$, respectively. In the following, we assume that $j \geq0$ and $k
\geq 0$ are coprime integers, i.e., they do not have a common divisor other than
1. For fixed aspect ratio $X$, it is obvious that the rectangles will attain maximum
volume $\upsilon_2(l, w)$ if they touch each other and line up precisely along their
short or long sides, cf.\ Fig.~\ref{fig:1}. Combining this and the periodicity of the
packing, it is straightforward to prove that ${\bf e}$ must be parallel to ${\bf
  R}_{jk}$ \cite{13}. In other words, for given $X$ maximum packing fractions will
always occur for ``rational'' orientations $\omega=\arctan(k/j)$ of the
rectangles. Maximal packings for specific $X$ can be constructed using the concept of
{\em lattice lines\/}. The line $L^{(0)}_{jk}$ through the origin is defined by the
lattice vector ${\bf R}_{jk}=(j,k)$. The distance of adjacent lattice sites on
$L^{(0)}_{jk}$ equals $l_{jk} = \sqrt{j^2+k^2}$. The square lattice can be decomposed
into a set of parallel lattice lines $L^{(\sigma)}_{jk}$, $\sigma=0$, $\pm1$,
$\ldots$, of distance $w_{jk}$, where $w_{jk}\,l_{jk}=\upsilon_0=1$ (cf.\ the dashed
lines in Fig.~\ref{fig:1}). Choosing $l=l_{jk}$ and $w=w_{jk}$, i.e., $X=X^{\rm
  max}_{jk} =l_{jk}/w_{jk}=j^2 + k^2$ leads to a perfect tiling with $\varphi_{\rm
  max} =\varphi_2 (X_{jk}^{\rm max})=l_{jk}\,w_{jk} =1$, for all coprime pairs ($j$,
$k$), cf.\ the maxima at $\varphi_2=1$ in the lower part of the main panel of
Fig.~\ref{fig:2}. The pairs $(j,k)$ can be ordered such that $X^{\rm max}_{\nu-1}
<X^{\rm max}_{\nu}$ where $X^{\rm max}_{\nu} = X^{\rm max}_{j_{\nu}k_{\nu}}$,
$\nu=0$, $1$, $2$, $\ldots$.  Since ${\bf e}$ must be parallel to ${\bf R}_{jk}$, the
maximum packing for $X <X^{\rm max}_{jk}$ and $X > X^{\rm max}_{jk}$ is obtained by
decreasing $l$ below $l_{jk}$ keeping $w=w_{jk}$, and decreasing $w$ below $w_{jk}$
keeping $l=l_{jk}$, respectively (see Fig.~\ref{fig:1}). Consequently,
\begin{eqnarray} \label{eq2}
\varphi_2(X)
=\left\{
\begin{array}{ll}
w_{j_{\nu}k_{\nu}}^2 X ,\quad   X^{\rm min}_{\nu-1}\leq  X \leq X^{\rm max}_{\nu}\\
l_{j_{\nu}k_{\nu}}^2/X ,\quad   X^{\rm max}_{\nu}\leq  X \leq X^{\rm min}_{\nu}
\end{array}
\right..
\end{eqnarray}
The positions $X^{\rm min}_{\nu}= l_{j_{\nu}k_{\nu}}l_{j_{\nu+1}k_{\nu+1}}$ follow
from the matching condition $(l_{j_{\nu}k_{\nu}})^2/X^{\rm min}_{\nu} =
(w_{j_{\nu+1}k_{\nu+1}})^2 X^{\rm min}_{\nu}$ and $l_{j_{\nu+1}k_{\nu+1}}
w_{j_{\nu+1}k_{\nu+1}} =1$.  $\varphi_2(X)$ is shown in Fig.~\ref{fig:2}, together
with the optimal orientation $\omega(X)$ in the inset.

%\begin{figure}
%\includegraphics[width=\linewidth]{Fig_3b}
%\caption{(Color online) Orientation angle $\omega$ as a function of $ X $:
%  Rectangles (blue line) ellipses (red line). The fat dots correspond to the position
%  of the maxima of $ \varphi_2(X) $ for ellipses.}
%\label{fig:3}
%\end{figure}

\begin{figure}
\includegraphics[width=\linewidth]{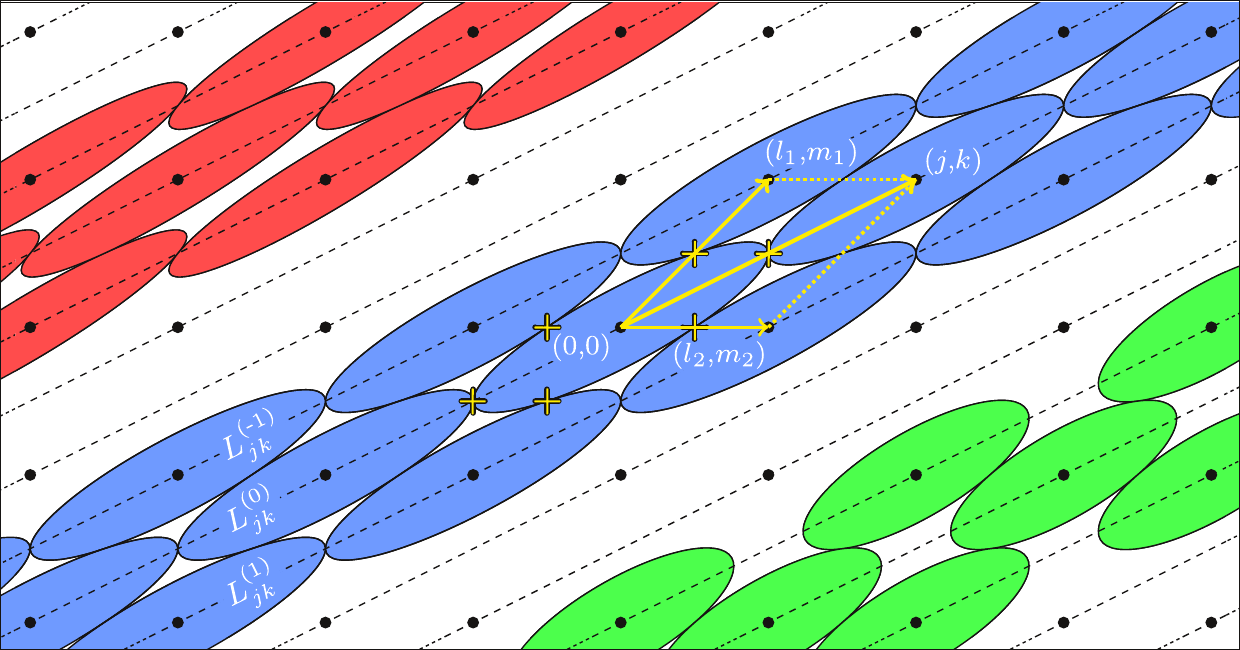}
\caption{(Color online) Maximum packing configurations of ellipses with aspect
  ratios $X=3$ (green, bottom), $X_{21}^{\text{max}}=\sqrt{29+8\sqrt{13}}/\sqrt{3}
  \cong 4.4$ (blue, middle), and $X=6$ (red, top). Lattice lines for $
  (j,k)=(2,1) $ (dashed lines) and contact points (crosses).}
\label{fig:4}
\end{figure}

We now turn to the case of packing ellipses on the square lattice. A naive approach
would be to inscribe them into the rectangles considered above. The resulting packing
fraction of ellipses is then just $\pi/4$ that of the rectangles. In reality,
however, maximally packed ellipses do not, in general, touch each other ``head'' to
``tail'', nor are they oriented parallel to the lattice lines, cf.\
Fig.~\ref{fig:4}. In contrast to the highly degenerate case of packing rectangles
which touch along whole line segments, packings of general, smooth convex bodies are
characterized by $K$ contact points per body of which, due to inversion symmetry,
only $K/2$ are independent. The three parameters describing an ellipse (two half axes
and the orientation angle) are under-determined in the generic case of $K=4$ contact
points (resulting in $K/2=2$ equations), yielding a {\em continuum\/} of solutions as
a function of $X$. Non-generic is the case of $K=6$ contacts, leading to a {\em
  discrete\/} set of maxima in $\varphi_2(X)$. For this situation, put one ellipse at
the origin, such that the sites of the other ellipses are at $\pm (l_i, m_i)$, $i=1$,
$2$ and $\pm(j,k)= \pm (l_1 + l_2, m_1 + m_2)$, cf.\ Fig.~\ref{fig:4}. Note that $0
\leq l_i \leq j$, $0 \leq m_i \leq k$. Then, the three corresponding contact vectors
${\bf c}_i=\frac{1}{2} (l_i, m_i)$, $i=1$, $2$ and ${\bf c}_3 =\frac{1}{2} (j,k)={\bf
  c}_1 + {\bf c_2}$ uniquely determine the three coefficients $a$, $b$, and $c$ in
the ellipse equation $ax^2 + 2bxy + cy^2=1$. This allows to determine the lengths of
the half axes and thus the aspect ratio to be
\begin{eqnarray} \label{eq6}
  X^{\rm max}_{jk} & = & (\alpha_++\sqrt{\alpha_-^2+\alpha_0^2})/\sqrt{3},\\
  \alpha_i & = & {\bf c}_1 \cdot {\boldsymbol \varepsilon}_i {\bf c}_1 +
  {\bf c}_2 \cdot {\boldsymbol \varepsilon}_i {\bf c}_2 +
  {\bf c}_1 \cdot {\boldsymbol \varepsilon}_i {\bf c}_2,\nonumber
\end{eqnarray}
where ${\boldsymbol\varepsilon}_0 = \begin{pmatrix} 0 & 1 \\ 1 & 0 \end{pmatrix}$ and
${\boldsymbol\varepsilon}_\pm = \begin{pmatrix} 1 & 0 \\ 0 & \pm
  1 \end{pmatrix}$. The corresponding packing fraction is $\varphi_{\rm max} =
\varphi_2 (X_{jk}^{\rm max})= \pi /\sqrt{12}$ for {\it all} $(j,k)$, identical to the
packing fraction of hcp disks. In fact, each such maximal ellipse packing can be
continuously deformed via an affine transformation into a packing of disks. On
increasing (decreasing) $X$ from $X^{\rm max}_{jk}$, the shortest (longest) contact
vector disappears, and the remaining four contacts allow to determine the
coefficients $a$, $b$ and $c$, and therefore $\varphi_2(X)$ and $\omega(X)$ in
between the maxima $X^{\rm max}_{jk}$ as a function of $X$ in a closed-form
expression.

The result for $\varphi_2(X)$ and $\omega(X)$ is displayed in Fig.~\ref{fig:2}. Since
$\varphi_2(1/X) = \varphi_2(X)$, only the regime $X\ge 1$ is shown. The maximum
packing fraction is singular at $X^{\rm max}_\nu$ for all $\nu$. The orientation
$\omega (X)$ is discontinuous at those $X_\nu^{\rm min}$ at which $\varphi'_2(X)$ is
discontinuous and at those $X_{jk}^\mathrm{max}$ which are degenerate, such as
$X_{51}^\mathrm{max} = X_{52}^\mathrm{max}$ (cf.\ Fig.~\ref{fig:2}). At these points,
there are two degenerate maximal packings with the same packing fraction and aspect
ratio, but different orientations. The global maximum value of $\varphi_{\rm max} =
1$ and $\varphi_{\rm max} = \pi/ \sqrt{12}$ for rectangles and ellipses,
respectively, is attained for an {\it infinite} number of packings, uniquely labeled
by $(j,k)$. From Fig.~\ref{fig:2} it appears plausible that
$\footnotesize\lim\limits_{X \rightarrow \infty} \varphi_2 (X)=\varphi_{\rm max}$,
which indeed can be proved \cite{13}.

The relation of the contact points can be understood from a number-theoretical point
of view. Note that the centers $(l_i, m_i)$, $i=1$, $2$ of two ellipses touching the
central one at $(0,0)$ also define lattice lines $L^{(0)}_{l_i m_i}$ with direction
$(l_i, m_i)$. These are the directions closest to that of $L^{(0)}_{jk}$ provided
that $l_i$ and $m_i$ are coprime and $0\leq l_i \leq j$, $0 \leq m_i \leq k$. In
mathematical terms, this means that $m_i/l_i$, $i=1$, $2$, are given by the best
principal and best intermediate rational approximant \cite{12} of $ k/j $. They
follow from the finite continued fraction expansion of $ k/j $,
\begin{equation} \label{eq7}
k/j=a_0+1/[a_1 + 1 / [a_2 + \cdots + 1/[a_{n-1} + 1/a_n] \cdots ]],
\end{equation}
where $a_i$, $i=1,\ldots,n$ ($ a_n\geq 2 $) are positive integers that are uniquely
determined by $k/j$.  Then, it is $l_1 = s_{n-1}$, $m_1 = r_{n-1}$, where the best
{\em principal\/} approximant $r_{n-1}/s_{n-1}$ follows from Eq.~\eqref{eq7} for
$a_n=\infty$, and $l_2 = s_{n,a_n-1}$, $m_2 = r_{n,a_n-1}$ follows analogously from
the best {\em intermediate\/} approximant $r_{n,a_n-1}/s_{n,a_n-1}$ obtained from
Eq.~\eqref{eq7} replacing $a_n$ by $a_n-1$. Since coding problems are strongly linked
to number theory \cite{3}, these results also promise insight into the connection
between packing and coding problems.

\begin{figure}
\includegraphics[width=\linewidth]{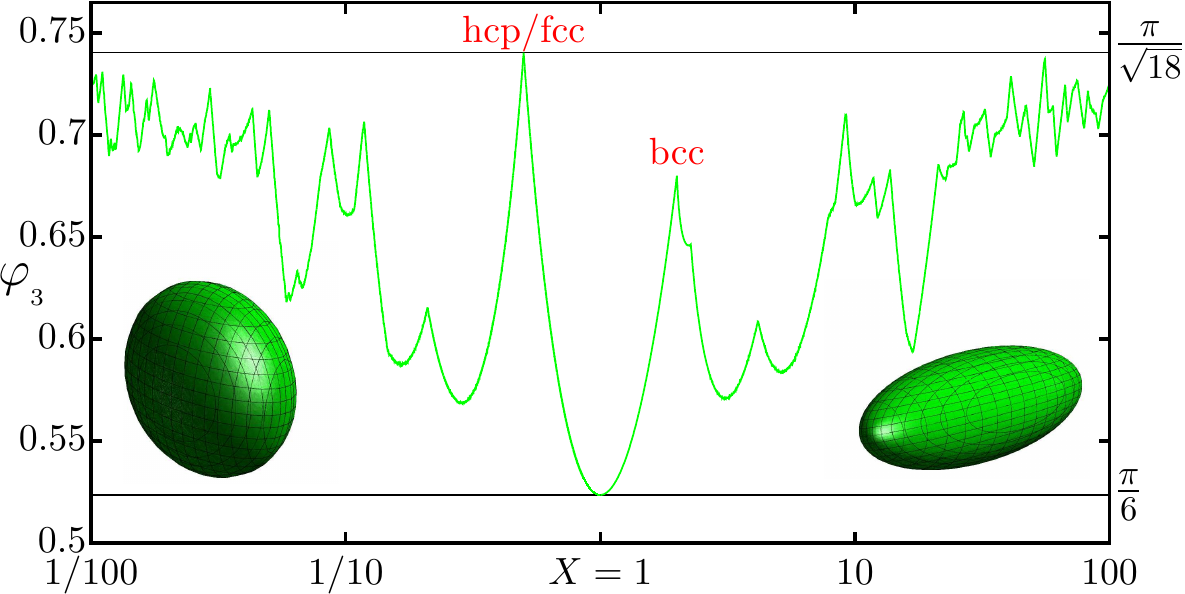}
\caption{(Color online) Maximum packing fraction for biaxial ellipsoids on an sc
  lattice on a logarithmic $X$ scale.}
\label{fig:5}
\end{figure}

Finally, we have investigated ellipsoids of revolution on a simple cubic
lattice. Analytically, it is possible to proceed in a similar fashion as for the
ellipses. The resulting eighth-order polynomial in $w^2$ can only be solved
numerically, however, and the intermediate expressions are rather
unwieldy. Therefore, we instead determined $\varphi_3(X)$ numerically by a
downhill-simplex minimization algorithm, the result of which is shown in
Fig.~\ref{fig:5}; it agrees with that determined earlier in Ref.~\cite{ricker} and,
as expected, shows continuity, too. Similar to the results in 2d, the derivative
$\varphi'_3(X)$ seems to be discontinuous at a series of maxima at $X_\nu^{\rm
  max}$. It appears to be discontinuous at some, but not all, minima $X_\nu^{\rm
  min}$. The symmetry between $1/X$ and $X$ valid in 2d, however, is lost. Most
strikingly, the global maximum $\varphi_{\rm max} = \pi/ \sqrt{18}$ for ellipsoids
appears to be attained only for the single packing fraction $X^{\rm max}_{-1} = 1/2$,
whereas in 2d there was a countable infinity of degenerate maxima. This maximum
corresponds to the highly non-generic case of each ellipsoid touching 12
neighbors. Consequently, an affine transformation can be applied to map this pattern
to closest packing of hard spheres resulting in an fcc or hcp lattice. A second
prominent maximum occurs at $X^{\rm max}_1=2$, where 8 contact points occur. The
corresponding transformed hard sphere packing yields a bcc lattice. It can be shown
that $\varphi_3(X) \geq\varphi_{\rm min} = \pi/6=\varphi_3(X=1)$, i.e., the packing
fraction of hard spheres on an sc lattice is a lower bound for $\varphi_3(X)$
\cite{13}. From the numerical results in Fig.~\ref{fig:5} we conjecture that,
analogous to the 2d case, $\varphi_3(X) \rightarrow \varphi_{\rm max}=\pi/\sqrt{18}$
for $X\rightarrow\infty$ and $X\rightarrow 0$, respectively.

To conclude, the maximum packing fraction $\varphi_d(X)$ of parallely aligned convex
objects ${\cal K}$ characterized by a single aspect ratio exhibits \emph{universal}
features that appear to be independent of ${\cal K}$, the underlying Bravais lattice
$\Lambda$, and even its dimension $d$. In particular, $\varphi_d(X)$ has been very
generally proved to be continuous. In fact, this proof can even be extended to the
case of convex bodies characterized by an arbitrary number of aspect
ratios. Furthermore, as shown for the case of rectangles and ellipses on the square
lattice as well as for biaxial ellipsoids on the sc lattice, there is an infinite
number of local maxima and minima at which $\varphi_d(X)$ is singular. The
singularities at the minima and at certain, degenerate maxima (see Fig.~\ref{fig:2}
for the case of ellipses) are correlated to the discontinuities in the orientation of
${\cal K}$. For the studied cases, we find that $\varphi_d(X)$ converges to its
global maximum for $X\rightarrow\infty$ as well as for $X\rightarrow 0$. While we
were only able to prove this rigorously for the case of rectangles and ellipses, we
believe that this property holds far more generally, implying that convex hard
objects, on average, pack much better if they become more oblate or prolate.

On the other hand, there are also significant differences between the systems studied
in two and three dimensions. For rectangles and ellipses, the global maximum packing
fraction $\varphi_\mathrm{max} = 1$ (rectangles) and $\varphi_\mathrm{max} =
\pi/\sqrt{12}$ (ellipses) is attained for an {\em infinite\/} number of discrete
aspect ratios $X_{jk}^\mathrm{max}$, uniquely labeled by pairs $(j,k)$ of coprime
integers.  On the contrary, for symmetric ellipsoids with $1/100\le X\le 100$,
$\varphi_3(X)$ takes its maximal height $\varphi_{\rm max}=\pi/\sqrt{18}$ at the {\it
  single} value $X^{\rm max}_{-1}=1/2$, only. This qualitative difference can be
understood as follows. Consider, for instance, a $d$-dimensional symmetric ellipsoid
which depends on $d+1$ parameters. In a packing, $K$ contacts lead to $K/2$
equations. In the generic case of $K/2 = d$, the system is under-determined and
$\varphi_d$ can be found as a function of the aspect ratio $X$. For the non-generic
case $K/2=d+1$, there is always a solution corresponding to the local maxima of
$\varphi_d(X)$ at $X^\mathrm{max}_\nu$. It appears likely that the competing point
symmetries of ${\cal K}$ and $\Lambda$ are responsible for the non-equal heights of
these maxima for $d=3$ and $K=8$. It is conceivable that this extra frustration might
be relieved by considering convex hard objects characterized by {\it three} length
scales, possibly leading again to an infinity of equal-height maxima. The even more
non-generic situation $K/2>d+1$ as realized, e.g., in the global maximum
$\varphi_\mathrm{max}=\pi/\sqrt{18}$ for our 3d ellipsoids with $K=12$, corresponds
to an {\em over-determined\/} set of equations such that, at most, only very few
solutions can be expected. It is worthwhile to point out that our results for
packing on fixed lattices should be closely related to ``continuum´´ packing with a
fixed number of contacts since the latter involves geometric frustration as well.

Of course, it might be a challenge to study packings of
the considered type on different lattices. Even richer behavior is expected on
weakening the condition of parallel alignment, paving the way for the occurrence of
superdense packings in analogy to those recently found for ellipsoids in the 3d
continuum \cite{10,11}.

M.W.\ acknowledges funding by the DFG under contract No. WE4425/1-1 (Emmy Noether
Program).

\end{document}